\documentstyle[prl,aps,epsf]{revtex} \def\narrowtext{} \tighten \twocolumn
\input epsf.sty 
\begin{document}
\draft
 
\title{Quasiparticle Liquid in the Highly Overdoped Bi$_{2}$Sr$_{2}$CaCu$_{2}$O$_{8+\delta}$}
\author{
        Z. Yusof,$^1$ B.O. Wells,$^1$
        T. Valla,$^2$ A.V. Fedorov,$^{2}$\cite{AF} P.D. Johnson,$^2$
        Q. Li,$^3$
	  C. Kendziora,$^4$
	  Sha Jian,$^5$ and D.G. Hinks,$^5$
        }
\address{
         (1) Department of Physics, University of Connecticut, 2152 Hillside Road U-46, Storrs, 		 CT 06269-3046.\\
         (2) Physics Department, Bldg. 510B, Brookhaven National Laboratory, Upton, NY 11973-		 5000.\\
         (3) Division of Materials Sciences, Brookhaven National Laboratory, Upton, NY 11973-		 5000.\\
	   (4) Naval Research Laboratory, Washington DC 20375.\\
	   (5) Materials Sciences Division, Argonne National Laboratory, Argonne IL 60439.\\
         }
\address{%
\begin{minipage}[t]{6.0in}
\begin{abstract}
We present results from the study of a highly overdoped (OD) Bi$_{2}$Sr$_{2}$CaCu$_{2}$O$_{8+\delta}$ with a $T_{c}=51$K using high resolution angle-resolved photoemission spectroscopy. The temperature dependent spectra near the ($\pi,0$) point show the presence of the sharp peak well above $T_{c}$. From the nodal direction, we make comparison of the self-energy with the optimally doped and underdoped cuprates, and the Mo(110) surface state. We show that this OD cuprate appears to have properties that approach that of the Mo. Further analysis shows that the OD has a more $k$-independent lineshape at the Fermi surface than the lower-doped cuprates. This allows for a realistic comparison of the nodal lifetime values to the experimental resistivity measurements via Boltzmann transport formulation. All these observations point to the validity of the quasiparticle picture for the OD even in the normal state within a certain energy and momentum range.
\typeout{polish abstract}
\end{abstract}
\end{minipage}}

\maketitle
\narrowtext

The question of whether the Fermi Liquid (FL)\cite{Nozieres} model is a valid description for the high-$T_{c}$ superconductors has been one of the central issues in condensed matter. For the optimally doped (OP), and certainly the underdoped (UD) cuprates, there are many results which indicate that the FL picture may not be valid\cite{Valla1,Anderson} especially in the normal state. It is speculated that as doping increases beyond the optimal value, the cuprate may become a FL.\cite{Batlogg} However, experimental evidence from the highly overdoped (OD) regime showing FL properties is scant. Moreover, the FL term has become imprecise in recent literature. In the strictest definition of a FL, transport properties and the self-energy, $\Sigma$, of the single-particle excitations or quasiparticles (QP) are proportional to the square of binding energy, $\omega$, and temperature, $T$, reflecting the electron-electron interactions. For regular metals, this is not the case for most $T$ and $\omega$ because the electron-phonon interaction dominates the scattering rate and the resistivity. At the other extreme, the term FL has been used to describe any material that has well defined single-particle excitation regardless of the nature of the QP scattering or transport. In this Letter, we describe in detail the nature of $\Sigma$ of the highly OD cuprate Bi$_{2}$Sr$_{2}$CaCu$_{2}$O$_{8+\delta}$ (Bi2212). We find a state where there are well-defined single-particle excitations but the details do not match either a strict definition of a FL nor does $\Sigma$ appear to be dominated by the electron-phonon interaction as found in normal metals. We use the term ``quasiparticle liquid'' (QPL) as a way to describe such a system that has: (i) well-defined single-particle excitations around the Fermi surface, and (ii) excitations that govern the transport properties even if these properties do not have the classic $\omega^{2}$ and $T^{2}$ behavior or electron-phonon-like behavior.
 
Angle-resolved photoemission spectroscopy (ARPES) is, in principle, the best method to determine the nature of the QP in a 2D solid. The presence of a sharp peak in the spectra potentially indicates the presence of a long-lived QP. Indeed, this has been used to argue for the presence of QP for OP and UD cuprates below $T_{c}$\cite{Kaminski1} However, the existence of a sharp peak is not in itself proof of the existence of a FL. In the normal state of the OP and UD, there may be a sharp peak but only in the nodal direction. In the superconducting state, the well-defined sharp peak appears around the Fermi surface but only for a very short range of $\omega$ and $k$ and appears to be unrelated to transport measurements. In fact, for most $\omega$ and $k$ values, the spectral function blows up into a broad, ill-defined excitation. Analysis of the peak in the nodal direction for OP cuprates reveals that $\Sigma$ is unlike that for a typical metal.\cite{Valla1} In model 2D metals, the extracted $\Sigma$ values are amenable to analyses in terms of traditional scattering mechanisms.\cite{Valla2-Mo} More exotic phenomenologies have been invoked to understand similar data for the OP.\cite{Valla1} Our approach is to not only present experimental data and analysis for highly OD Bi2212 but also to compare the results to the lower-doped compounds and to a metallic 2D surface state of Mo(110). This allows us to determine crucial differences amongst cuprates of different dopings, and between the cuprates and other real materials.

Single crystals of Bi2212 were synthesized using the floating-zone technique\cite{Miyakawa1} that yielded optimal $T_{c}$ onset of 95 K. Overdoping is achieved by annealing the samples in oxygen at 400 C for 5 days, yielding $T_{c}\sim $ 51K.\cite{Kendziora} The samples were kept in the cells under oxygen pressure until they were mounted in the vacuum chamber. The crystals were cleaved in situ at $T$ below 150 K under vacuum with a base pressure of $8\times 10^{-11}$ Torr. Crystal orientation was determined using low-energy electron diffraction (LEED). All data shown here were obtained within 12 hours after cleaving to minimize variation in doping content of the exposed surface with time. $T_{c}$ of the samples were determined to be $\sim $51K $\pm $ 5K using a SQUID magnetometer. All ARPES measurments were performed at beamline U13UB at the NSLS with photon energy of 22 eV and a Scienta SES 200 hemispherical analyzer that simultaneously collects a large energy (0.5 to 1 eV) and angular (12$^{0}$) window. This allows for the simultaneous collection of the photoemission intensity as a function of binding energy (energy distribution curve or EDC) and momentum (momentum distribution curve or MDC) along a particular direction in the BZ.\cite{Valla1} The resulting energy resolution is $\sim $10 meV with an angular resolution of better than 0.2$^{0}$.

Fig. 1a shows the $T$ dependence of the EDCs near the M-point ($\Gamma $M direction) of the BZ. The spectra show a well-defined sharp peak along with a broader hump that persist well into the normal state, a behavior that was also seen in a similar OD material.\cite{Rast} In OP and UD cuprates the sharp peak disappears around $T_{c}$.\cite{Fedorov} A recent report\cite{Feng} shows the sharp peak disappearing at slightly above $T_{c}$ on lightly OD crystals and claims the peak intensity is a measure of the superfluid density. This cannot be the case for our more OD samples since the sharp peak persists to temperatures well above $T_{c}$ where there is no superfluid.\cite{Rast} The only change in lineshape as $T$ goes above $T_{c}$ is a shift in the leading edge to signify the closing of the gap, as shown in Fig 1b. The gap size estimated from this shift is 10 meV, which is consistent with that observed in tunneling measurements on similar crystals.\cite{Lutfi-comm}. 

A visual demonstration of the different energy ranges over which the ARPES excitation is well defined has been shown in Ref. \cite{Johnson} for the Bi2212 cuprates and Ref. \cite{Valla2-Mo} for Mo(110). When compared to the lower-doped cuprates, the OD ARPES spectra show a larger range of $\omega$ and $k$ over which the spectral peak is well defined. $k_{F}$ for the OD in this direction is measured to be 0.39A$^{-1}$ or 0.34($\pi ,\pi $), compared to 0.446A$^{-1}$ or 0.391($\pi ,\pi $) for OP Bi2212\cite{Valla1}, indicating a shift of the Fermi surface towards the $\Gamma $ point. In contrast to these samples, the Mo shows well defined peaks over the entire band. Clearly the $\omega$ range over which the ARPES spectra is well defined is much greater for the OD sample versus either the UD or OP, but still less than that for the Mo.

The width $\Delta k$ of the MDC spectra in the $\Gamma$Y region is used to obtain Im$\Sigma$ using the relation $\hbar v_{k} \Delta k \sim 2$Im$\Sigma$,\cite{Valla1} where $v_{k}$ is the non-interacting band velocity.\cite{Johnson} Fig. 2 shows Im$\Sigma$ along with similar data for the OP and the Mo. None of these data sets show Im$\Sigma$ varying as $\omega^{2}$ for small energies. As noted previously, the Mo data can be broken into contributions from three scattering mechanisms: electron-electron, electron-phonon, and defect scattering.\cite{Valla2-Mo} The phonon mechanism dominates in the vacinity of $E_{F}$. Unlike Mo, 2Im$\Sigma$ for the OD and the OP Bi2212\cite{Valla1} have a predominantly linear dependence on $\omega$. For the OP sample, the slope of Im$\Sigma$ versus $\omega$ is much greater and follows a characteristic Marginal Fermi Liquid (MFL) behavior where Im$\Sigma \sim$ max($\omega,T$).\cite{Varma} Such an identification is less clear in the OD sample, but appears possible. Certainly the OD does not seem to be easily analyzed in terms of the typical scattering processes in metals. Overall, the peak broadening that gives Im$\Sigma$ for the OD sample lies in between the behavior of the OP and the Mo. 

In both the normal and superconducting state, $\Sigma$ of the OD cuprate is much less $k$ dependent than for the lesser-doped compounds. This is manifested in the ARPES lineshape at a given $\omega$ at different parts of the BZ. Fig. 3a and b compare the OD EDC's from the two in-plane symmetry directions at several binding energies, both below and above $T_{c}$. There is a remarkable similarity in the lineshape along $\Gamma $M and $\Gamma $Y. The main difference is that the remnant broad hump seen for $\omega _{p}\sim $ 12 meV is only present in the $\Gamma $M spectra. This hump becomes less pronounced at higher $\omega$ where the spectra look more similar to each other. We compare this to the spectra of OP Bi2212 (Fig. 3d) where there is a dramatic difference between the spectra from the two directions. This is especially true above $T_{c}$ where there is no well defined peak in the $\Gamma $M spectrum. The difference in lineshapes is even stronger in UD Bi2212. We conclude that well-defined ARPES peaks occur over a much greater $k$ range in the normal state of the OD compound versus the OP or UD material.\cite{Valla3} In particular $\Sigma$ seems to be $k$ dependent in the UD and OP samples but largely $k$ independent in the OD material. It could be argued that with increasing hole doping, the influence of the antiferromagnetic ``environment'' in the system becomes weaker. This is consistent with the weaker coupling strength with increasing doping observed in the nodal direction.\cite{Johnson}

In conventional metals, transport is governed by QP that scatter due to interactions with other excitations. The resistivity $\rho$ in these materials are related to the lifetime or the mean free path via Boltzmann transport theory that involves an integral over the Fermi surface.\cite{Ashcroft} Although the lifetimes of ARPES QP, and thus ARPES peak widths, are not identical to that obtained in transport measurements, they should be roughly governed by the same scattering processes. This peak width is directly measured in ARPES via $\Delta k$, the width of the MDC peak at $E_{F}$. For the OD samples, we have QP that are well defined over the whole Fermi surface and scattering rates that are largely $k$-independent as seen in Fig. 3. Thus, it is realistic to consider that in the OD case, the Boltzmann transport formulation would produce a meaningful comparison between the lifetime and the measured in-plane resistivity $\rho_{ab}$, and that the lifetime or mean free path obtained in the nodal direction can be taken as the average value. Such consideration results in the general relationship where $\rho_{ab}$ is directly proportional to $\Delta k$. Figure 4 shows a comparison of the ARPES $\Delta k$ data with $\rho_{ab}$.\cite{Kendziora} The resistivity data comes from a differently-prepared crystal that is overdoped to have roughly the same $T_{c}$, the best comparison available. We find that both $\rho_{ab}$ and $\Delta k$ have the same functional form. Note that $\Delta k$ appears to be insensitive to the superconducting transition. The similarity in the temperature dependence of the normal state ARPES $\Delta k(E=E_{F})$ and $\rho_{ab}$ is a strong evidence that both transport and ARPES are dominated by quasiparticles with similar scattering interactions.

Recent ARPES studies of OD cuprates have shown the possiblity of two separate bands arising out of the bilayer splitting.\cite{Feng2,Chuang} While our data (not shown) may indicate evidence for a two-component Fermi surface, the specific labeling of the hump feature in the EDC spectra near ($\pi,0$) as the bonding band\cite{Feng2} is inconsistent with our data in terms of the $T$ dependence and trends of relative peak intensities with doping as shown in Fig 1. We note that the mere presence of separate bilayer split bands has little impact on the conclusions of this paper. In fact, the idea of interplane coupling appearing in the OD samples is understandable in terms of the results reported here. Since the interplane coupling is dominated by states near the M-points of the BZ,\cite{Ioffe} the presence of more well-defined QP in those regions for the OD would increase the interplane coupling that possibly leads to the bilayer splitting. Furthermore, the well-defined QP would enhance the coherent interlayer tunneling, resulting in a greater $c$-axis conductivity. This has been shown to be the case where the $c$-axis resistivity decreases and appears to be metallic with increasing doping.\cite{Yan} 

Taken as a whole, there are several elements of the ARPES data which indicate that the OD Bi2212 may be considered a QPL in the normal state, though the primary interactions are considerably different than those in a prototypical metal. The key elements arguing for the QPL nature of the normal state of the OD sample are: (i) the presence of well-defined peaks in the ARPES spectra over a wide range of $T$, $\omega$ and $k$, along with well defined peaks for the entire Fermi surface, (ii) $\Sigma$ that is largely $k$ independent, unlike the UD and OP samples, and (iii) $\rho_{ab}$ that is proportional to the mean free path measured in ARPES as is expected from simple Boltzman transport. The lower-doped cuprates do not seem to be well described as a QPL in the normal state. The fundamental interactions between the electrons and other excitations is similar for all of the cuprates, and not like that for a typical metal. The strongest evidence for this is the linear $\omega$ dependence of Im$\Sigma$ as shown in Fig. 3 which do not reflect contributions from several different scattering processes and no sign of contributions from the electron-phonon interaction. One alternative scenario involves some form of coupling to the magnetic modes observed in inelastic neutron scattering.\cite{Bourges} The overall changes in the ARPES $\Sigma$ as a function of $T$ and doping can be matched to the magnetic behavior.\cite{Johnson} However, regardless of the details of the interactions, there is a clear difference between the OD and OP cuprates in both the presence of QP and their relation to transport properties.

Work supported in part by Dept. of Energy under contract number DE-AC02-98CH10886, DE-FG02-00ER45801, DOE-BES W-31-109-ENG-38, and in part by the New Energy and Industrial Technology Development Organization, Australia. The authors acknowledge discussions with John Zasadzinski,  Liam Coffey, Phil Allen, Donglai Feng, and Bob Gooding.

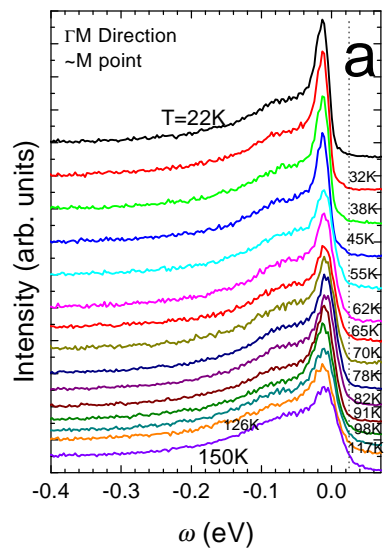
FIG. 1. (color) EDCs for OD Bi2212 ($T_{c}=51$K) near the M-point (see inset) at various temperatures. The data in Fig. 1a have been shifted vertically for clarity. Six representative spectra are shown unshifted in Fig. 1b.

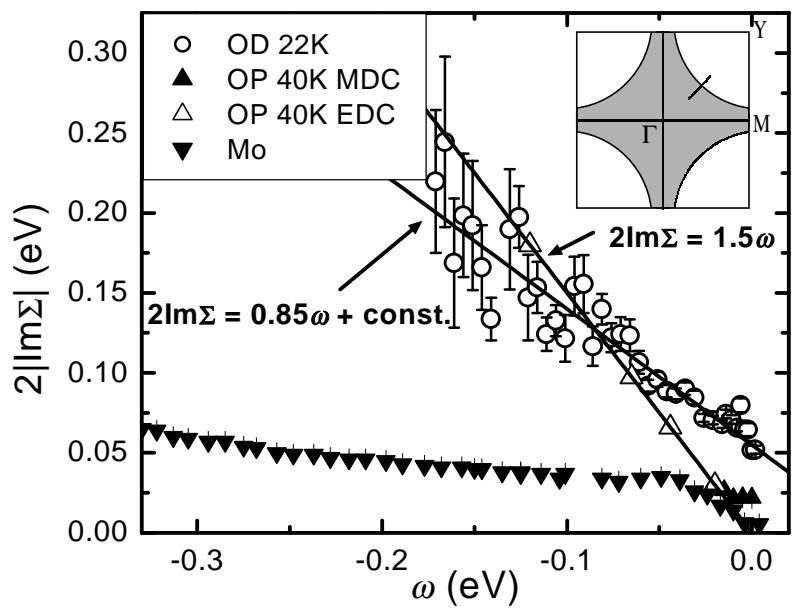
FIG. 2. 2Im$\Sigma$ in the $\Gamma$Y direction (see inset) for OD ($T_{C}=51$K), OP ($T_{C}=91$K),\cite{Valla1} and Mo(110) surface state.\cite{Valla2-Mo} The values shown in the legend are the temperatures where the ARPES spectra were obtained. 2Im$\Sigma$ for the OD was obtained exclusively from the MDC spectra, whereas the OP values were obtained from both MDC and EDC. The scattering due to impurities has been subtracted from the Mo data. The straight lines are guides to the eye.

FIG. 3. (color) EDC lineshapes for the OD (a,b, and c) and OP (d) taken at the M-point and the nodal direction. Each data have been normalized to the intensity of the sharp peak, except for the normal state of OP where it has not been normalized. $\omega_{p}$ is the binding energy of the peak intensity.

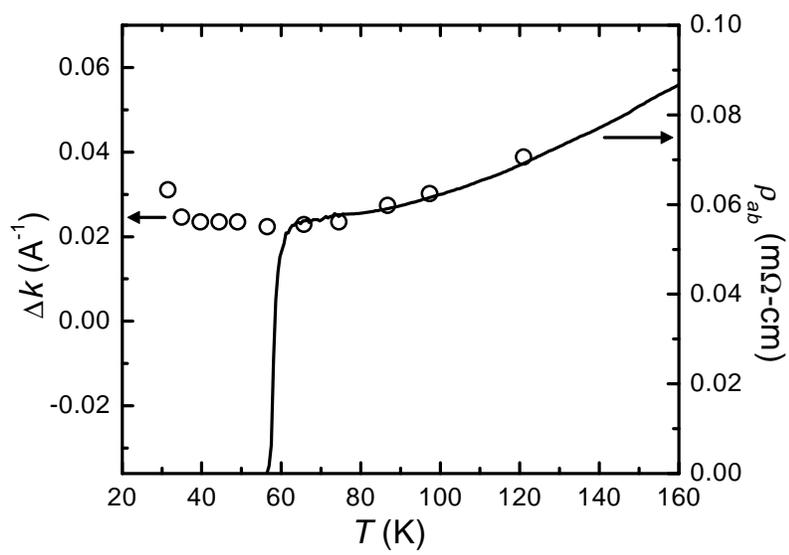
FIG. 4. A comparison of the ARPES $\Delta k$ (open circles) and experimental $\rho_{ab}$ (solid line) from OD Bi2212 with $T_{c}=58$K.\cite{Kendziora}.


\begin{references}

\bibitem[\dag]{AF} Present address: Dept. of Physics, University of Colorado, Boulder, CO 80309-	0390 and Advanced Light Source, Lawrence Berkeley National Laboratory, Berkeley, CA 94720.

\bibitem{Nozieres} P. Nozieres, \textit{Theory of Interacting Fermi Systems} (Addison-Wesley, 	Reading, 1964).

\bibitem{Valla1} T. Valla \textit{et al}., Science \textbf{285}, 2110 (1999).

\bibitem{Anderson} P.W. Anderson, \textit{Theory of Superconductivity in the High-$T_{c}$ 	Cuprates} (Princeton University Press, Princeton NJ, 1997).

\bibitem{Batlogg} B. Batlogg and C.M. Varma, Phys. World \textbf{13}, 2 (2000).

\bibitem{Kaminski1} A. Kaminski \textit{et al}., Phys. Rev. Lett. \textbf{84}, 1788 (2000).

\bibitem{Valla2-Mo}  T. Valla \textit{et al}., Phys. Rev. Lett. \textbf{83}, 2085 (1999).

\bibitem{Miyakawa1}  N. Miyakawa \textit{et al}., Phys. Rev. Lett. \textbf{80}, 157 (1998).

\bibitem{Kendziora} C. Kendziora \textit{et al}., Physica C 257, 74 (1996).

\bibitem{Rast}  S. Rast \textit{et al}., Europhys. Lett. \textbf{51}, 103 (2000).

\bibitem{Fedorov}  A.V. Fedorov \textit{et al}., Phys. Rev. Lett. \textbf{82}, 2179 (1999).

\bibitem{Feng} D.L. Feng \textit{et al}., Science \textbf{289}, 277 (2000).

\bibitem{Lutfi-comm}  J.F. Zasadzinski \textit{et al}., cond-mat/0102475.

\bibitem{Johnson} P.D. Johnson \textit{et al}., cond-mat/0102260.

\bibitem{Varma} C.M. Varma \textit{et al}., Phys. Rev. Lett. \textbf{63}, 1996 (1989).

\bibitem{Valla3} T. Valla \textit{et al}., Phys. Rev. Lett. \textbf{85}, 828 (2000).

\bibitem{Ashcroft} N.W. Ashcroft and N.D. Mermin, \textit{Solid State Physics} (Saunders, 1976).

\bibitem{Feng2} D.L. Feng \textit{et al}., cond-mat/0102385.

\bibitem{Chuang} Y.-D. Chuang \textit{et al}., cond-mat/0102386.

\bibitem{Ioffe} L.B. Ioffe and A.J. Millis, Science \textbf{285}, 1241 (1999); A.I. Liechtenstein 	\textit{et al}., Phys. Rev. B \textbf{54} (1996).

\bibitem{Yan} Y.F. Yan \textit{et al}., Phys. Rev. B \textbf{52}, 751 (1995); T. Watanebe 	\textit{et al}., Phys. Rev. Lett. \textbf{79}, 2113 (1997).

\bibitem{Bourges} P. Bourges, \textit{The Gap Symmetry and Fluctuations in High Temperature 	Superconductors}, ed. J. Bok, G. Deutscher, D. Pavuna, and S.A. Wolf (Plenum Press, 1998).

\end{references}
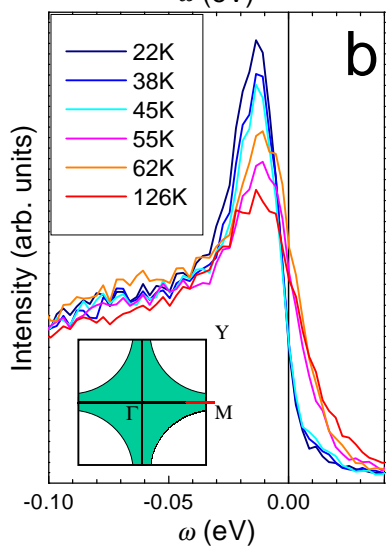
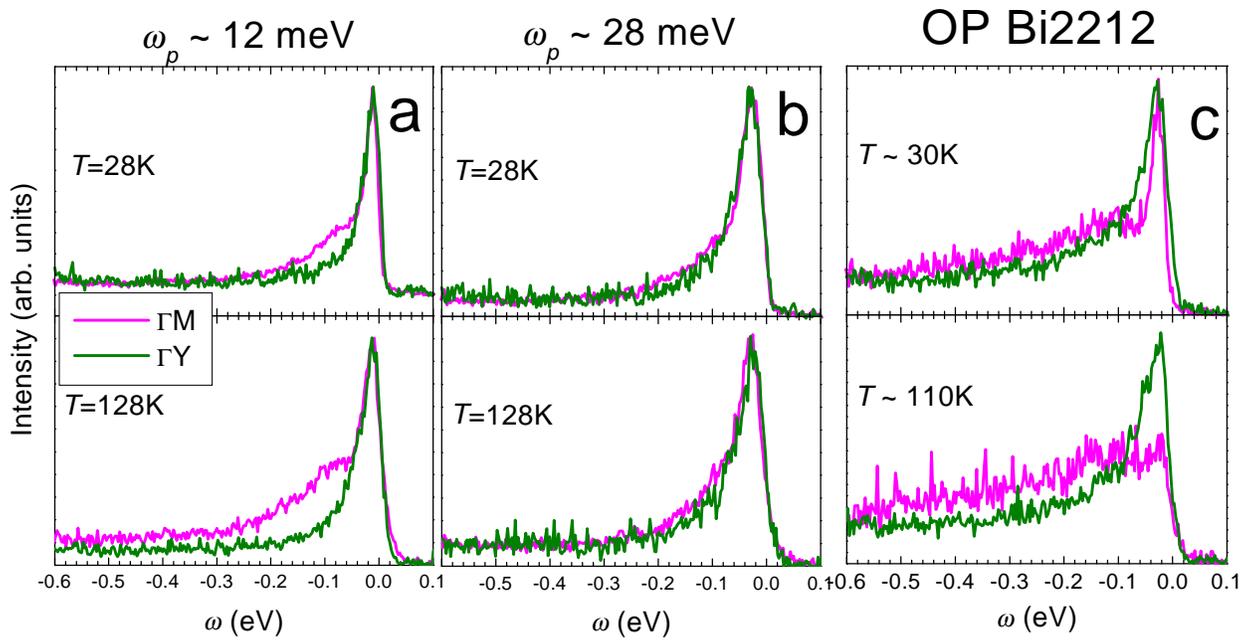
\end{document}